\documentclass[10pt]{article}
\usepackage{amsmath}
\usepackage{amssymb}
\usepackage{hyperref}
\usepackage{graphicx}
\usepackage{epstopdf}
\usepackage{cite}
\usepackage{color}
\usepackage{filecontents}
\usepackage[labelfont=bf,labelsep=period,justification=raggedright]{caption}
\makeatletter
\renewcommand{\@biblabel}[1]{\quad#1.}
\makeatother

\date{}
\begin{document}
\begin{flushleft}

{\Large \textbf{The Dynamics of Collaborative Knowledge Production} }
\\
Lingfei Wu$^{1,\ast}$, Jacopo A. Baggio$^{1,2}$, Marco A. Janssen$^{1,3}$
\\
\bf{1} Center for Behavior, Institutions and the Environment, Arizona State University, Tempe, AZ 85281, U.S.
\\
\bf{2} Department of Environment and Society, Utah State University, 84322, Logan UT, U.S.
\\
\bf{3} School of Sustainability, Arizona State University, Tempe, AZ 85281, U.S.
\\
$\ast$ E-mail: Lingfei.Wu@asu.edu
\end{flushleft}

\section*{Abstract}

Online communities are becoming increasingly important as platforms for large-scale human cooperation. These communities allow users seeking and sharing professional skills to solve problems collaboratively. To investigate how users cooperate to complete a large number of knowledge-producing tasks, we analyze StackExchange, one of the largest question and answer systems in the world. We construct attention networks to model the growth of 110 communities in the StackExchange system and quantify individual answering strategies using the linking dynamics of attention networks. We identify two types of users taking different strategies. One strategy (type A) aims at performing maintenance by doing simple tasks, while the other strategy (type B) aims investing time in doing challenging tasks. We find that the number of type A needs to be twice as big as type B users for a sustainable growth of communities. 




\section*{Introduction}
Humans are unique in their ability to create public goods in non-repeated situations with non-kin. In larger groups cooperation is more difficult due to the higher temptation to free ride on the voluntary contributions of others \cite{olson2009logic}. Nevertheless humans are able to create public goods with thousands and even millions of unrelated individuals. For example, there are an increasing number of online communities where participants put in time and effort to make voluntary contributions such as street maps \cite{zook2010volunteered}, software \cite{schweik2012internet}, encyclopedic information \cite{jemielniak2014common}, protein folding \cite{khatib2011algorithm}, and language translation \cite{von2013duolingo}.

	Online communities are natural experiments that give us an opportunity to test possible mechanisms that explain cooperation in large groups. Controlled online experiments show that if participants can choose group members higher levels of cooperation can be derived \cite{rand2011dynamic}. This suggests that assortment is a sufficient condition to derive cooperation in large groups. However, such experiments have a duration of about an hour in which participants are all simultaneously online and are recruited with the promise of monetary payments. Whether this scales up to large groups over longer periods of time is an open question.
	
	We will demonstrate in this paper that assortment is not sufficient to derive high levels of contributions in online collaboration \cite{henrich2004foundations}. Our analysis shows that at least two different types of strategies of making voluntary contributions are needed to sustain an online community over a longer period of time. One strategy (type A) aims at performing maintenance by doing simple  tasks, while the other strategy (type B) aims investing time in doing challenging tasks. We cannot measure the motivations for those two strategies, but we hypothesize that the first may related to reputation in the broader community, and the second to intrinsic motivations and reputation among peers. 
	
	For our empirical analysis we investigate the answering records of nearly three million users over a period of six years from 110 online communities. We find that type A users are important in cutting down the median waiting time for answers, while type B users help increase the acceptance rate of answers. The comparison of overall size across the studied communities suggests that a ratio equals 3:2 between type A and B users is preferred for bigger communities. We use these empirical finding to build an ``attention" network model in which we formalize the strategies depicted above as linking dynamics. In an attention network model the nodes are questions and the edges are successive answering activities connecting two questions. This model allows us to analyze the effect of individual answering behavior on the growth of communities. Our analysis not only supports the existence of a trade-off between the two types of users, but also predicts what will happen when the ratio in a community deviates from the optimal ratio detected. We predict that a community containing too many type A users lacks high quality answers, thus making it difficult to attract new questions continuously. On the contrary, a community composed of too many type B users has many high quality questions, but it will attract more new questions than it can handle. In sum, a balance between the two types of users is necessary for the sustainable growth of communities. At the end of the paper, we select three communities to illustrate the predictions of our attention network model, including ``math.stackexchange.com" (which has an optimal ratio), ``astronomy.stackexchange.com" (which contains too many type A users), and ``electronics.stackexchange.com" (which contains too many type B users).

\section*{Mixing Strategies and the Sustained Growth of Communities}

Figure \ref{strategies} depicts the profiles of the two types of users. Type A users prefer the easier, newer questions and have an answer acceptance rate higher than type B users, who have a tendency to answer the more difficult, older questions, across all expertise levels. The ANOVA tests on all the four variables between the two groups are significant (Figure \ref{strategies} B $\sim$ C).

It is natural to ask whether the ratio between type A and B users has an effect on the overall performance of communities. Two important indicators concerning the performance of Q\&A communities are the median waiting time for answers and the overall acceptance rate of answers \cite{mamykina2011design,bosu2013building}. 
The waiting time is defined as the time interval between a question being posted and a satisfying answer being accepted. Median is used instead of mean, because the distribution of waiting time comprises a few extremely large values that will lead to biased results if using the mean value \cite{mamykina2011design}. The accepted answer rate is the fraction of the questions with an accepted answer over the total population of questions. A good community is expected to have a high acceptance rate and a short median waiting time \cite{mamykina2011design,bosu2013building}. 

\begin{figure*}[ht]
\begin{center}
\centerline{\includegraphics[width=0.8\textwidth]{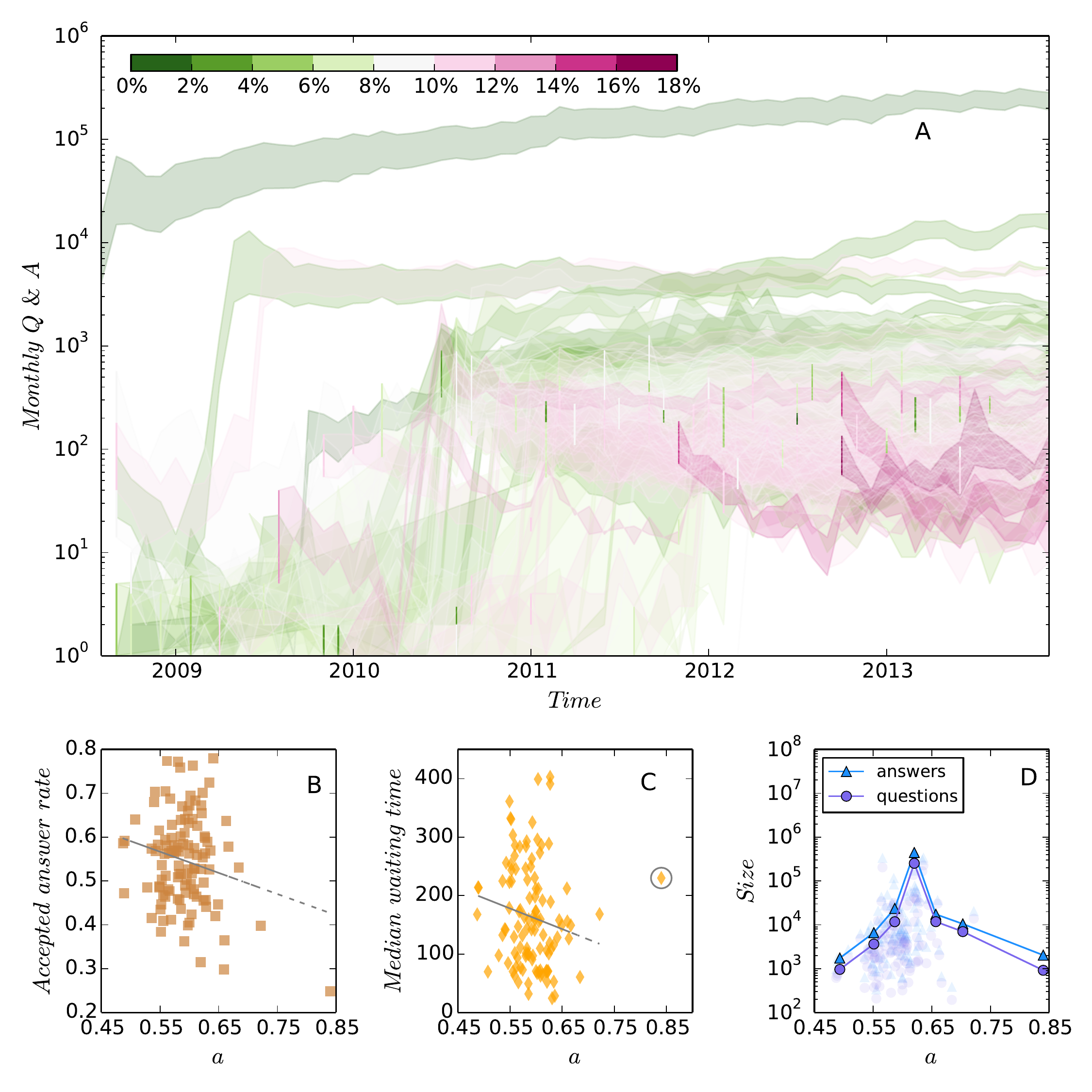}}
\caption{The effect of the fraction of type A users on the performance of communities. We find that both of accepted answer rate (B) and median waiting time (C) decrease with the increase of type A users. As a good community is expected to have a high accepted answer rate and a short waiting time of answers \cite{mamykina2011design,bosu2013building}, our finding suggests that a balance between the two types of users should be carefully chosen in order to optimize the performance. In (B) the regression coefficient is $-0.48$ and the p-value is $0.029$. In (C) after removing the outlier (the data point in the gray circle) the regression coefficient is $-352.89$ and the p-value is $0.107$. Figure (D) suggests that the maximum size of communities is archived when $a\approx0.63$. Figure (A) shows the monthly increased number of questions and answers of 110 sites, in which the upper bounds of bands show the number of answers and the lower bounds show the number of questions. These bands are plotted in different colors to shown their derivation from the optimal ratio $a=63\%$. It is observed that the larger deviation from the optimal ratio is related with more unsustainable growth. See online version for color display.}
\label{sus}
\end{center}
\end{figure*}

In Figure \ref{sus} we calculate the fraction of type A users $a$ ($a\in [0,1]$) in each of the 110 communities and plot the two discussed indicators against $a$. It turns out that type A users contribute to the waiting time and type B users contribute to the accepted answer rate. Therefore, a balance between these two types of users should be carefully chosen in order to optimize the community performance. In Figure \ref{sus}(C) we find that the maximum community size (``stackoverflow.com") is achieved when $a$ approximates $0.63$, i.e., the ratio of type A users to type B users is 3:2. In Figure \ref{sus}(A) we plot the growth curves of all communities and color them based on their derivation from the optimal $a$. This figure shows clearly how either too many type A users or two many type B users leads to an unsustainable growth rate.

\section*{From Answering Strategies to Linking Dynamics}

We have shown a co-occurrence between an optimal ratio and the maximum community size. However, why a deviation from the optimal ratio is related to small communities is still unclear. Therefore, we use a network model to formalize our assumptions of individual answering strategies and to analyze the consequences of different ratios.

We define attention networks to represent Q\&A communities, in which nodes are questions and edges are the sequential answering activities of users. In attention networks, an answering strategy based on the number of the existing answers to questions can be interpreted as a degree-based rule of link increases. As type A users prefer easy questions (low-degree nodes) and type B users favor difficult questions (high-degree nodes), when these two types of users respond to a new question, they carry links from very different nodes. Therefore, we can simplify the model by assuming that type A strategy corresponds to the rule of ``preferential attachment" \cite{barabasi1999emergence}, in which the rich get richer, and type B strategy corresponds to the reversed process of ``preferential attachment" \cite{sevim2006effects}, in which the attractiveness of a node decreases with its degree. The reversed ``preferential attachment" process is usually used to describe resource-based competition between nodes in flow networks such as food webs \cite{dunne2002food}, power grids\cite{amaral2000classes}, and the airport network \cite{guimera2004modeling}. For example, in food webs an outbound edge transport resources from a ``prey" node to a ``predator" node. If several predators are fed on the same prey, then the supplied resources have to be split and shared, thus decreasing the attractiveness of the prey node. We argue that this effect also exist in attention networks in which questions are competing for the limited attention of users. 

We use $f$ and $1-f$ ($f\in$[0,1]) to represent the probability of observing type A and B strategies, respectively (note that $f$ is different from the empirical value of $a$ mentioned in the last section, as $a$ is not the fraction of activities but the fraction of users), and quantify the probability $p(k)$ of a new question being connected to a pre-existing similar question of degree $k$ as

\begin{equation}
\label{eq.linkpro}
p(k) = f\frac{\frac{1}{k}}{\sum \frac{1}{k}} + (1-f) \frac{k}{\sum k}.
\end{equation}

As introduced, in two extreme cases this model degenerates to the ``preferential attachment" model ($f=0$) and the ``reversed preferential attachment" model ($f=1$) , respectively. Using the master equation technique \cite{dorogovtsev2000structure} we derive that the tail of the degree distribution will converge to 
\begin{equation}
\label{eq.degreedist}
p_k \sim k^{-\alpha} = k ^{-\frac{3-f}{1-f}},
\end{equation}
in which the power exponent $\alpha$ has a minimum value $3$ and always increases with $f$ (see SI for the details). We find that for a majority of communities the empirical value of $\alpha$ lies in the scope of [3,5] (Figure \ref{scalingsB}M), supporting our derivation. As a larger power-law exponent implies higher equality in resource distribution \cite{newman2005power}, our model suggests that type A strategy equalizes the allocation of attention (edges) among nodes and increases the chance of a new question being answered. 

Besides degree distribution, the discussed linking rules also explain several scaling relationships observed in the growth dynamics of attention networks as presented by Figure \ref{scalingsB} and mentioned in \cite{leskovec2005graphs,wu2011acceleratingb}. Users are more likely to post questions when they search the Web and find that a similar question has obtained many answers but their concerns have not been fully addressed. As a consequence, a new question is more likely to be added to the network if the existing similar questions have more answers. To include this process in our model, we consider node replication as the main driving force underlying network growth and allow high-degree nodes to generate more new nodes. Considering the node-matching probability $p(k)$ given by Eq.\ref{eq.linkpro}, we can calculate that the expected attractiveness of a single nodes is
\begin{equation}
\label{eq.expectdegree}
E(k) = \sum_{k=1}^{k_{max}}kp(k)\sim N^{\frac{1-f}{2}}.
\end{equation}
Therefore, the expected number of new nodes generated by an existing node is $E(k)N^g$, in which we use $N^g$ to model the effect of network size. By summing $E(k)N^g$ over all nodes in the network we obtain the total number of new nodes as $\Delta N = E(k)N^{g+1}$. 
Substituting this condition into Eq. \ref{eq.expectdegree} we derive the scaling relationship between the number of new and old nodes
\begin{equation}
\label{eq.newnodes}
\Delta N \sim N^\eta=N^{\frac{3-f}{2}+g}.
\end{equation}
Note that if an old node generates many new nodes, then there will be a stronger competition between these new nodes for edges. As a result, the cost of linking to an existing node is proportional to its degree \cite{sevim2006effects}. Meanwhile, it is reasonable to assume that new questions cannot obtain an infinite number of answers but have a limited ``quota" that approximates constant $C$. Putting these two conditions together, we derive the expected number of links obtained by a new node as $\Delta m = CN^h/E(k)$, in which $N^h$ is the effect of network size. 
Using the conclusion of Eq. \ref{eq.expectdegree} we have
\begin{equation}
\label{eq.avelinks}
\Delta m \sim N^\delta = N^{\frac{f-1}{2}+h}.
\end{equation}
From Eq. \ref{eq.newnodes} and Eq. \ref{eq.avelinks} we can derive the scaling relationship between the number of new edges and new nodes:
\begin{equation}
\label{eq.newlinks}
\Delta M = \Delta m \Delta N \sim \Delta N^\gamma = \Delta N^{\frac{\delta}{\eta}+1} = \Delta N^{\frac{2(h+g+1)}{3-f+2g}}
\end{equation}

To summarize, the above analysis explains why a mixture of different strategies is crucial to the sustainable growth of communities. On one hand, Eq. \ref{eq.avelinks} suggests that a community should have more type A users to maintain the number of answers per question (larger $\delta$); on the other hand, Eq. \ref{eq.newnodes} suggests that a community should have more type B users to attract new questions (larger $\eta$). As a balance, Eq. \ref{eq.newlinks} predicts that an optimmal  fraction of type A users, $f = 3+2g$, is preferred in order to maximize the value of $\gamma$, i.e., to match the growth of answers with the growth of questions. We argue that $f$ as the fraction of behavior can be viewed as the multiplication between two variables, the fraction of users $a$ and answering frequency distribution $w$. As $w$ is always positive, $f$ changes in the same direction as $a$. Therefore the derived optimimal value of $f$ implies that there is also an optimal value of $a$, which is consistent with our empirical observation.

\section*{Examples of Successful and Less Successful Communities}

\begin{figure*}[ht]
\begin{center}
\centerline{\includegraphics[width=0.9\textwidth]{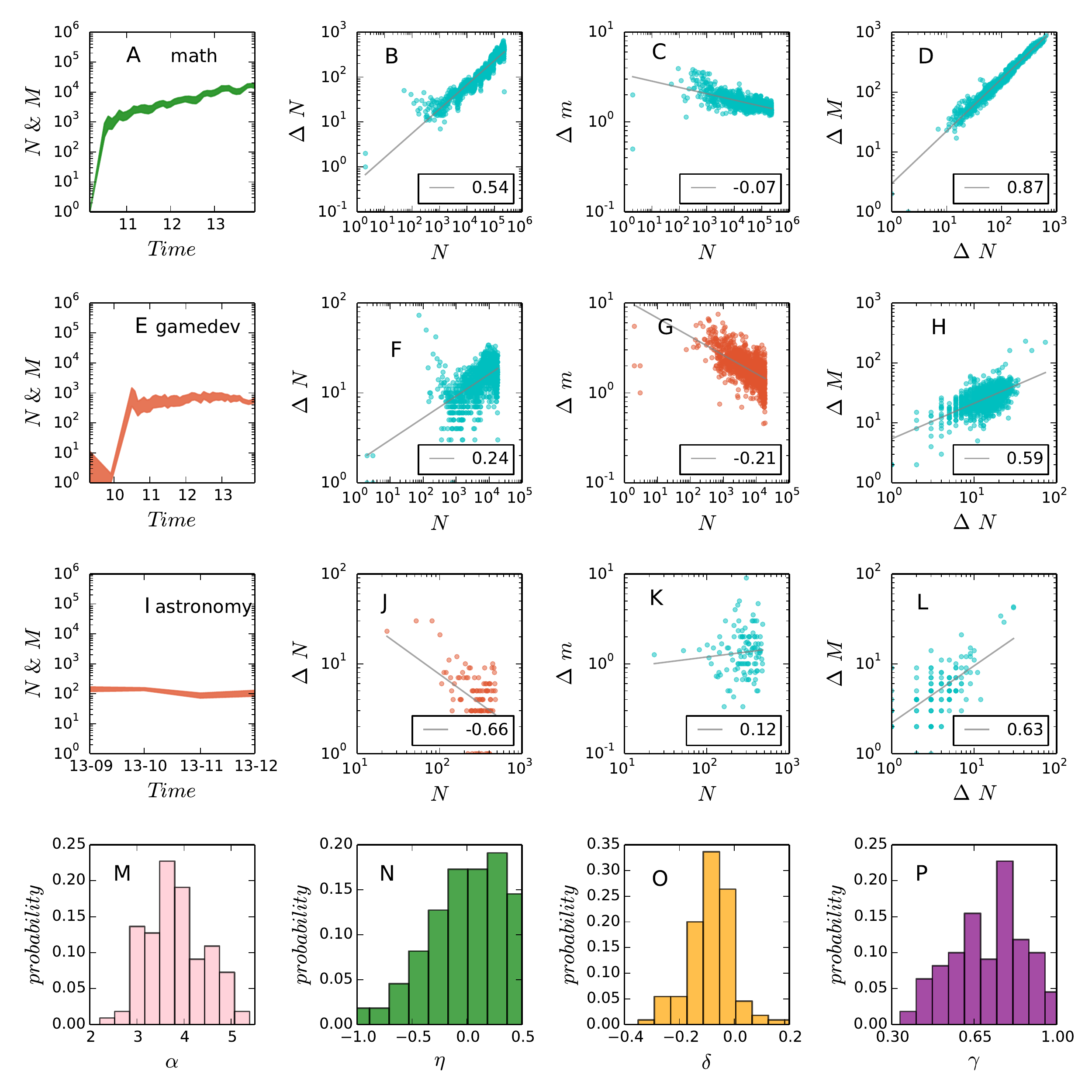}}
\caption{The scaling properties of communities. We select ``math.stackexchange.com" ($a = 0.63$, the first row) as the example for sustainable growth and ``gamedev.stackexchange.com" ($a = 0.62$, the second row) and ``astronomy.stackexchange.com" ($a = 0.67$, the third row) as the examples for unsustainable growth. In (A), (E), and (I) we plot the monthly increased number of questions (the lower bounds of the bands) and answers (the upper bounds of the bands) against time. The rest of figures in the first three rows demonstrate the scaling behaviors of communities. In particular, the second column corresponds to Eq. \ref{eq.newnodes}, the third column corresponds to Eq. \ref{eq.avelinks}, and the fourth column corresponds to Eq. \ref{eq.newlinks}. In these figures the OLS regression (gray) lines in logarithmic axes are plotted to demonstrate the estimated scaling exponents. We find that as predicted by our model,  in ``gamedev.stackexchange.com" new questions increase so fast (F) that they cannot be answered on time (G); while in ``astronomy.stackexchange.com" questions are answered quickly (K) but the increase of new questions is very slow (J). The last row gives the distributions of the four scaling parameters across the 110 communities. The mean values of the parameters are 3.7 ($\alpha$), -0.04 ($\eta$), -0.08 ($\delta$), and 0.71 ($\gamma$). See online version for color display.
}
\label{scalingsB}
\end{center}
\end{figure*}

Three communities are selected to compare the consequences of different ratios between type A and B users (Figure \ref{scalingsB}), including a community for math questions (math.stackexchange.com, or MATH in short), a community for questions about astronomy (astronomy.stackexchange.com, or ASTR), and a community for questions about game development (gamedev.stackexchange.com, or GAME). As given by the first column in Figure \ref{scalingsB}, ASTR and GAME are not as successful as MATH in maintaining a sustained growth curved, and this be explained by our model.  

The fraction of type A users in MATH is approximately $0.63$, which is equal to the optimal value. In contrast, ASTR has more type A users ($a = 0.67$). According to our model, questions in ASTR will be responded to efficiently, but there will be a slow growth of new questions. This prediction is confirmed by Figure \ref{scalingsB}k, which shows that the average number of answers to new questions is increasing, and Figure \ref{scalingsB}J, which shows that the increase of new questions slows down as time goes on. Meanwhile, GAME has a few more type B users ($a = 0.62$) than the optimum. According to our model, in this community new questions will increase so fast that they cannot be answered quickly. The fast increase of new questions is evidenced in Figure \ref{scalingsB}F, and the shrinking budget of answers per question is observed in Figure \ref{scalingsB}G. It is interesting to note that ASTR is one of the most traditional scientific areas and GAME is a new, fast developing area due to the widespread use of smart phones. We argue that neither simply being classic nor being hot would naturally lead to sustained growth, instead, the sustained growth of a community comes from the careful balance between contributors of diverse preferences.

\section*{Conclusions and Discussions}

We look at online communities as natural experiments for collective action problems. Our results imply that assortment is not sufficient to derive high levels of contributions in massive collaboration. Instead, strategic diversity seems to be the key for sustainable online communities. In the Stack Exchange datasets, a mixing ratio of $3:2$ between two types of users is found to maximize the size of communities. Type A users have a tendency to answer easier, newer questions and type B users prefer to answer more difficult, older questions. We propose an attention network model to formalize the two answering strategies of users and to explain the existence of an optimal ratio. Our conclusion is that type A users contribute to the number of answers and type B users contribute to the quality of answers, thus both of them is crucial to the development of communities.  

Our work generalizes the models of Barabasi et al. \cite{barabasi1999emergence} and Sevim et al. \cite{sevim2006effects} to study large-scale cooperation in online communities. The present analysis on attention networks can also be applied to model a variety of other online collective behaviors such thread browsing \cite{wu2013decentralized}, photo tagging \cite{wu2011acceleratinga, cattuto2007semiotic, wu2013metabolism}, and news sharing \cite{wu2007novelty}. The current study also has limitations, which point out the future direction of research. For example, to obtain a simple math model we simplify the rich behaviors of users and only consider two extreme strategies. Meanwhile, in attention networks we naively assume that the ratio between the two types of users is a constant throughout the evolution of a community. In future studies we may consider a ratio that changes over time.

\section*{Materials and Methods}

\subsection*{Data sources}

Stack Exchange is a network of question and answer communities covering diverse topics 
in many different fields. We downloaded its database dump on January, 2014 from 
\url{https://archive.org/details/stackexchange},
which contains the log files of 110 communities.  
The smallest community \url{italian.stackexchange.com} was created in November, 2013 and 
has 374 users, 194 questions, and 387 answers in our data set. The largest site \url{stackoverflow.com} (SO) was 
created in July, 2008 and has 2,728,224 users, 6,474,687 questions and 11,540,788 answers in our data set.


\subsection*{Measuring Question Difficulty and User Expertise}

\begin{figure*}[ht]
\begin{center}
\centerline{\includegraphics[width=1.1\textwidth]{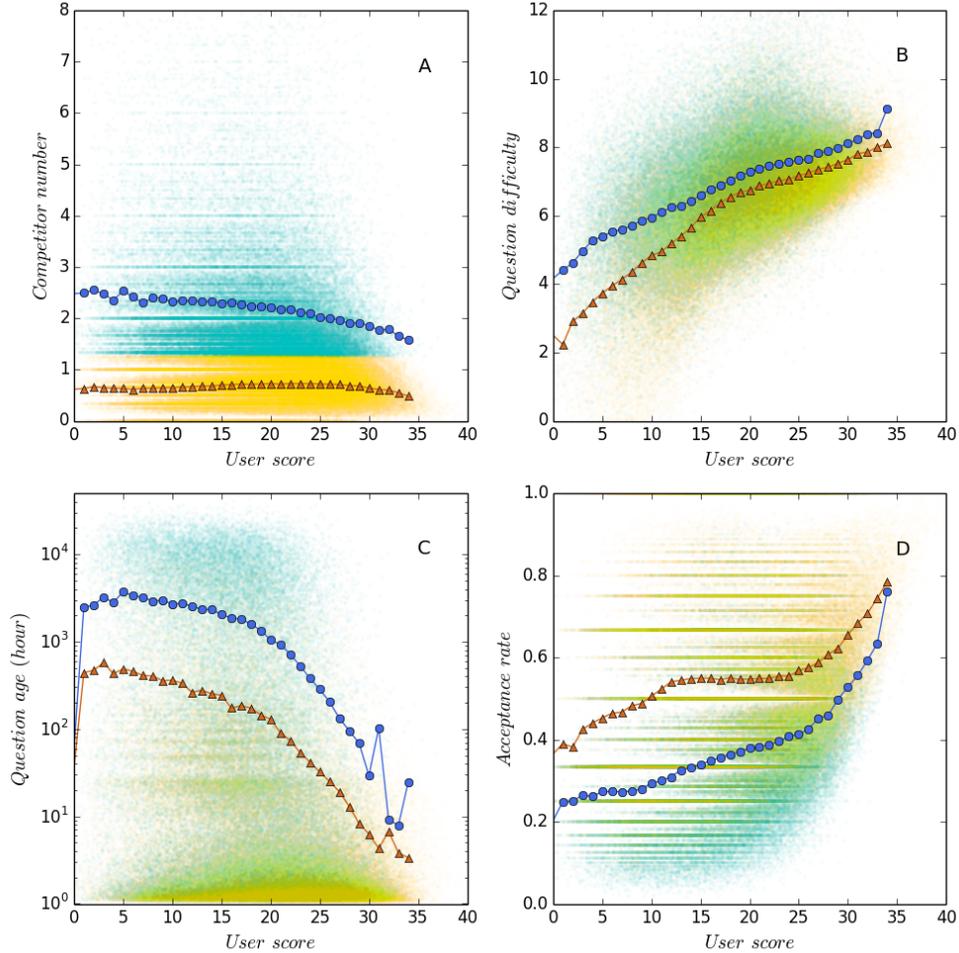}}
\caption{Two opposite answering strategies in the SO data set. Each data point shown in the background is a user. For each user, we use his expertise score calculated from the TrueSkill algorithm as the $x$-axis of the data point and use several user-level metrics as the $y$-axis. These metrics include the average number of competitors in answering questions (A), the average difficulty of answered questions (B), the average age of answered questions (C), and the overall acceptance rate of answers (D). We use the mean value of the first variable, the average number of competitors, as a threshold and separate users into two groups, type B users and type A users. The fraction of type A users is $0.63$ in the SO dataset. In all figures, we plot the linear-binned data of the two groups for comparison \cite{newman2005power}. We find that on average, type A users (triangles) tend to choose newer and easier questions and have higher answer acceptance rates than type B users (circles). See online version for color display.
}
\label{strategies}
\end{center}
\end{figure*}

We use the number of answers as a proxy for the ``perceived difficulty" of questions \cite{hanrahan2012modeling} in order to divide users according to their difficulty preferences. We firstly count the number of existing answers $q_{ij}$ to a question $j$ when a user $i$ responds to it. Then we average this number over all $m$ questions answered by a user to derive his/her average level of difficulty preference $\frac{1}{m}\sum_{j=1}^{m} q_{ij}$. After that, we use the grand mean of difficulty preferences in a community containing $n$ users, which is $\frac{1}{mn}\sum_{i=1}^{n}\sum_{j=1}^{m} q_{ij}$, as the threshold to separate type A users (whose preference is smaller than the threshold) from type B users (whose preference is greater than the threshold) (Figure \ref{strategies} A). To validate the difference between the two types of users, we compare the other two variables, including the average age of answered questions and the acceptance rate of submitted answers (Figure \ref{strategies} C $\sim$ D). 

The TrueSkill algorithm \cite{herbrich2006trueskill,liu2013question} was applied to validate the estimation on users' difficulty preferences. In the SO community we obtain the TrueSkill scores of 912,082 users and 3,771,021 questions (please see SI for the details), which represent the expertise levels of users and the ``real difficulty" of questions, respectively \cite{liu2013question}. We find that type B users are always answering more difficult questions than type A users at all expertise levels (Figure \ref{strategies} B). In our opinion this result validates our division of the two groups. Furthermore, we find that the 
TrueSkill scores of users are positively correlated with their reputation points in the log files (Pearson coefficient $\rho$ = 0.29,  p-value $<$ 0.001), justifying the credibility of TrueSkill scores.

\subsection*{Constructing Attention Networks}

The answering strategies of individual users can be understood from a network perspective, in which two questions are connected if they are answered sequentially by the same users (see SI for the details). From this perspective, Q\&A communities are growing networks with increasing nodes (questions) and links (answers). We call them ``attention networks" because they show the transportation of collective attention in solving problems. Attention networks translate answering strategies into linking dynamics; hence provide a quantitative, predictive model for us to explore the collective answering behavior of users. 

From empirical data we construct a growing attention network for each of the 110 communities. The network properties we are interested in include the cumulative number of nodes ($N$) and edges ($M$), the daily increments of nodes ($\Delta N$) and edges ($\Delta M$), and the number of links per node ($m = M/N$) and its daily increments ($\Delta m = \Delta M/ \Delta N$).

\section*{Author contributions}

L. W. and M.A.J. designed research; J.B. contributed new analysis ideas; L.W. analyzed data; L.W., J.B., and M.A.J. wrote the paper.

\section*{Acknowledgments}

We acknowledge financial support for this work from the National Science Foundation, grant number 1210856.

\bibliography{ref}

\begin{thebibliography}{10}
\providecommand{\url}[1]{\texttt{#1}}
\providecommand{\urlprefix}{URL }
\expandafter\ifx\csname urlstyle\endcsname\relax
  \providecommand{\doi}[1]{doi:\discretionary{}{}{}#1}\else
  \providecommand{\doi}{doi:\discretionary{}{}{}\begingroup
  \urlstyle{rm}\Url}\fi
\providecommand{\bibAnnoteFile}[1]{%
  \IfFileExists{#1}{\begin{quotation}\noindent\textsc{Key:} #1\\
  \textsc{Annotation:}\ \input{#1}\end{quotation}}{}}
\providecommand{\bibAnnote}[2]{%
  \begin{quotation}\noindent\textsc{Key:} #1\\
  \textsc{Annotation:}\ #2\end{quotation}}
\providecommand{\eprint}[2][]{\url{#2}}

\bibitem{olson2009logic}
Olson M, Olson M (2009) The logic of collective action: public goods and the
  theory of groups., volume 124.
\newblock Harvard University Press.
\bibAnnoteFile{olson2009logic}

\bibitem{zook2010volunteered}
Zook M, Graham M, Shelton T, Gorman S (2010) Volunteered geographic information
  and crowdsourcing disaster relief: a case study of the haitian earthquake.
\newblock World Medical \& Health Policy 2: 7--33.
\bibAnnoteFile{zook2010volunteered}

\bibitem{schweik2012internet}
Schweik CM, English RC (2012) Internet success: a study of open-source software
  commons.
\newblock MIT Press.
\bibAnnoteFile{schweik2012internet}

\bibitem{jemielniak2014common}
Jemielniak D (2014) Common Knowledge?: An Ethnography of Wikipedia.
\newblock Stanford University Press.
\bibAnnoteFile{jemielniak2014common}

\bibitem{khatib2011algorithm}
Khatib F, Cooper S, Tyka MD, Xu K, Makedon I, et~al. (2011) Algorithm discovery
  by protein folding game players.
\newblock Proceedings of the National Academy of Sciences 108: 18949--18953.
\bibAnnoteFile{khatib2011algorithm}

\bibitem{von2013duolingo}
von Ahn L (2013) Duolingo: learn a language for free while helping to translate
  the web.
\newblock In: Proceedings of the 2013 international conference on Intelligent
  user interfaces. ACM, pp. 1--2.
\bibAnnoteFile{von2013duolingo}

\bibitem{rand2011dynamic}
Rand DG, Arbesman S, Christakis NA (2011) Dynamic social networks promote
  cooperation in experiments with humans.
\newblock Proceedings of the National Academy of Sciences 108: 19193--19198.
\bibAnnoteFile{rand2011dynamic}

\bibitem{henrich2004foundations}
Henrich J, Boyd R, Bowles S, Camerer C, Fehr E, et~al. (2004) Foundations of
  human sociality: Economic experiments and ethnographic evidence from fifteen
  small-scale societies.
\newblock Oxford University Press.
\bibAnnoteFile{henrich2004foundations}

\bibitem{mamykina2011design}
Mamykina L, Manoim B, Mittal M, Hripcsak G, Hartmann B (2011) Design lessons
  from the fastest q\&a site in the west.
\newblock In: Proceedings of the SIGCHI conference on Human factors in
  computing systems. ACM, pp. 2857--2866.
\bibAnnoteFile{mamykina2011design}

\bibitem{bosu2013building}
Bosu A, Corley CS, Heaton D, Chatterji D, Carver JC, et~al. (2013) Building
  reputation in stackoverflow: an empirical investigation.
\newblock In: Proceedings of the 10th Working Conference on Mining Software
  Repositories. IEEE Press, pp. 89--92.
\bibAnnoteFile{bosu2013building}

\bibitem{barabasi1999emergence}
Barab{\'a}si AL, Albert R (1999) Emergence of scaling in random networks.
\newblock science 286: 509--512.
\bibAnnoteFile{barabasi1999emergence}

\bibitem{sevim2006effects}
Sevim V, Rikvold PA (2006) Effects of preference for attachment to low-degree
  nodes on the degree distributions of a growing directed network and a simple
  food-web model.
\newblock Physical Review E 73: 056115.
\bibAnnoteFile{sevim2006effects}

\bibitem{dunne2002food}
Dunne JA, Williams RJ, Martinez ND (2002) Food-web structure and network
  theory: the role of connectance and size.
\newblock Proceedings of the National Academy of Sciences 99: 12917--12922.
\bibAnnoteFile{dunne2002food}

\bibitem{amaral2000classes}
Amaral LAN, Scala A, Barthelemy M, Stanley HE (2000) Classes of small-world
  networks.
\newblock Proceedings of the National Academy of Sciences 97: 11149--11152.
\bibAnnoteFile{amaral2000classes}

\bibitem{guimera2004modeling}
Guimera R, Amaral LAN (2004) Modeling the world-wide airport network.
\newblock The European Physical Journal B-Condensed Matter and Complex Systems
  38: 381--385.
\bibAnnoteFile{guimera2004modeling}

\bibitem{dorogovtsev2000structure}
Dorogovtsev SN, Mendes JFF, Samukhin AN (2000) Structure of growing networks
  with preferential linking.
\newblock Physical Review Letters 85: 4633.
\bibAnnoteFile{dorogovtsev2000structure}

\bibitem{newman2005power}
Newman ME (2005) Power laws, pareto distributions and zipf's law.
\newblock Contemporary physics 46: 323--351.
\bibAnnoteFile{newman2005power}

\bibitem{leskovec2005graphs}
Leskovec J, Kleinberg J, Faloutsos C (2005) Graphs over time: densification
  laws, shrinking diameters and possible explanations.
\newblock In: Proceedings of the eleventh ACM SIGKDD international conference
  on Knowledge discovery in data mining. ACM, pp. 177--187.
\bibAnnoteFile{leskovec2005graphs}

\bibitem{wu2011acceleratingb}
Wu L, Zhang J (2011) Accelerating growth and size-dependent distribution of
  human online activities.
\newblock Physical Review E 84: 026113.
\bibAnnoteFile{wu2011acceleratingb}

\bibitem{wu2013decentralized}
Wu L, Zhang J (2013) The decentralized flow structure of clickstreams on the
  web.
\newblock The European Physical Journal B 86: 1--6.
\bibAnnoteFile{wu2013decentralized}

\bibitem{wu2011acceleratinga}
Wu L (2011) The accelerating growth of online tagging systems.
\newblock The European Physical Journal B-Condensed Matter and Complex Systems
  83: 283--287.
\bibAnnoteFile{wu2011acceleratinga}

\bibitem{cattuto2007semiotic}
Cattuto C, Loreto V, Pietronero L (2007) Semiotic dynamics and collaborative
  tagging.
\newblock Proceedings of the National Academy of Sciences 104: 1461--1464.
\bibAnnoteFile{cattuto2007semiotic}

\bibitem{wu2013metabolism}
Wu L, Zhang J, Min Z (2014) The metabolism and growth of web forums.
\newblock PloS one 8: e102646.
\bibAnnoteFile{wu2013metabolism}

\bibitem{wu2007novelty}
Wu F, Huberman BA (2007) Novelty and collective attention.
\newblock Proceedings of the National Academy of Sciences 104: 17599--17601.
\bibAnnoteFile{wu2007novelty}

\bibitem{hanrahan2012modeling}
Hanrahan BV, Convertino G, Nelson L (2012) Modeling problem difficulty and
  expertise in stackoverflow.
\newblock In: Proceedings of the ACM 2012 conference on Computer Supported
  Cooperative Work Companion. ACM, pp. 91--94.
\bibAnnoteFile{hanrahan2012modeling}

\bibitem{herbrich2006trueskill}
Herbrich R, Minka T, Graepel T (2006) Trueskill: A bayesian skill rating
  system.
\newblock In: Advances in Neural Information Processing Systems. pp. 569--576.
\bibAnnoteFile{herbrich2006trueskill}

\bibitem{liu2013question}
Liu J, Wang Q, Lin CY, Hon HW (2013) Question difficulty estimation in
  community question answering services.
\newblock In: EMNLP. pp. 85--90.
\bibAnnoteFile{liu2013question}

\end{thebibliography}

\end{document}